\documentclass[3p,times,fleqn]{article}
\usepackage[T1]{fontenc}
\usepackage[english]{babel} 
\usepackage[utf8]{inputenc} 
\usepackage{amssymb, amsmath, pxfonts} 
\usepackage{mathrsfs} 
\usepackage[normalem]{ulem} 
\usepackage{mathrsfs} 
\usepackage[top=3cm,left=3cm,right=2cm,bottom=2cm]{geometry} 
\usepackage{graphicx} 
\usepackage[usenames]{color} 
\usepackage{makeidx} 
\usepackage{indentfirst} 
\usepackage[footnotesize]{caption}

\usepackage{calligra}
\usepackage{color}
\usepackage{float}
\graphicspath{ {images/} }
\usepackage{multicol}

\begin{document}

\begin{center}
\Large{Thermodynamic States of the Mixed Spin 1/2 and Spin 1 Hexagonal Nanowire System Obtained from a Seven-Site Cluster Within an Improved Mean Field Approximation}
\end{center}

\vspace{1cm}

\begin{center}

R. G. B. Mendes $^{(a)}$,
F. C. Sá Barreto $^{(a)(b)}$,
J. P. Santos $^{(a)(c)}$

\vspace{0,5cm}

\textit{ ,(a) Departamento de Ci\^{e}ncias Naturais, Universidade Federal de S\~{a}o Jo\~{a}o del Rei, C.P. 110, CEP 36301-160, \\S\~ao Jo\~ao del Rei, MG, Brazil}

\vspace{0,5cm}

\textit{ (b) Emeritus Professor, Departamento de F\'{i}sica, Universidade Federal de Minas Gerais, C.P. 110, CEP 31270-901, \\ Belo Horizonte, MG, Brazil}

\vspace{0,5cm}

\textit{ (c) Departamento de Matem\'atica, Universidade Federal de S\~{a}o Jo\~{a}o del Rei, C.P. 110, CEP 36301-160, \\S\~ao Jo\~ao del Rei, MG, Brazil}

\end{center}

\vspace{1,7cm}

\section*{Abstract}
The mean field approximation results in the mixed spin 1/2 Ising model and spin 1  Blume-Capel model, in the hexagonal nanowire system, are obtained from the Bogoliubov inequality. The Gibbs free energy, magnetization and critical frontiers are obtained. Besides the stable branches of the order parameters, we obtain the metastable and unstable parts of these curves and also find phase transitions of the metastable branches of the order parameters. The classification of the stable, metastable and unstable states is made by comparing the free energy values of these states.\\

\textit{Keyword}: Gibbs free energy, Bogoliubov inequality, entropy, hexagonal nanowire, Blume-Capel, Ising, mixed-spin, metastable transition

\section*{Introduction}

Magnetic nanowire and nanotube, based in transition metals, have received high attention due their potential application in high density magnetic recording media \cite{ExpNanowire1, ExpNanowire2,ExpNanowire3}, in bio-technology \cite{ExpNanowireMEDICINAL, ExpNanowire4MEDICINAL} and light sensor \cite{EXpNwZnOSensor}. The favorable feature for the use of nanotubes and nanowire, in transition metals, is their higher surface effects when compared to the  bulk materials.

In magnetic nanowires and nanotube with multilayers, a special kind of structure is the structure with different spins value in each layer, called mixed-spin. In the study of the magnetism of these structures the effective field theory(EFT) \cite{MixedNwEFT, MixedNwEFT2, MixedNwEFT3, MixedNtEFT} and Monte Carlo simulation \cite{MixedNwEFTMonteCarlo, MixedNwMonteCarlo, MixedNwMonteCarlo2} are the main technical procedures used. Through these procedures the phase diagrams are found and behaviors like hysteresis and influence between core and shell spins are observed.

The Bogoliubov inequality \cite{J1, J2, J3} for the free energy functional is an inequality that gives rise to a variational principle of statistical mechanics. Choosing a trial hamiltonian $H_{o}$ to be a simple approximation for the true hamiltonian $H$ of some lattice-statistical model, it affords a rigorous upper bound for the true Gibbs free energy. Recently, the variational principle of Bogoliubov inequality and generalized mean fields in many-particle interacting systems were studied \cite{J7, J8}. Maksimov and Kuzemsky \cite{J9} studied the cluster generalization of the mean field approximation for the case of a two spin-cluster. The mean field approximation(MFA) was also studied on the Blume-Capel model in a five-site cluster on the diamond lattice from the Bogoliubov inequality \cite{J10}. The mean field approximation (MFA) presumes a virtual independence between the spins variables, and  the interactions acting in each variable spin like an \textit{effective field} caused by the nearest neighbours sites\cite{Weiss1,Weiss2}. If compared with \textit{Effective Field Theory}, which considers the autocorrelation between spins, the most advantage of the MFA is the possibility to describe better the system's symmetry by the use of a larger number of spins in the described cluster without much increase in mathematical complexity.

\begin{figure}[H]
\centering
  \includegraphics{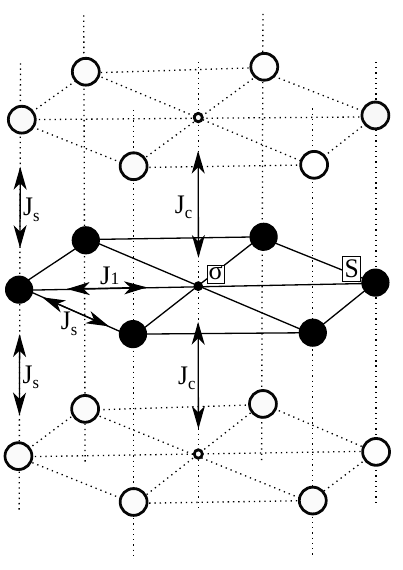}
  \captionof{figure}{Mixed magnetic nanowire with spin 1/2 Ising in the central column and spin 1 Blume-Capel in the external hexagonal layer.}
  \label{fig.1}
\end{figure}

The objective of this work is to present the magnetic behavior and the thermodynamics of the stable, metastable and unstable states of a magnetic nanowire, Fig- \ref{fig.1}. The $\sigma$ spins of the central column are $\sigma=1/2$ Ising type while the $S$ in external hexagonal are  $S=1$ Blume-Capel type. To describe the whole systems symmetry a hexagonal cluster was chosen. For this cluster, it was used the Bogoliubov inequality to find the Gibbs free energy $G$ by the mean field approximation. Through $G$ it was obtained the phase diagram of the first order, second order and metastable phase transitions, and the magnetizations and entropy values for stable, metastable and unstable states. 

In this work it is not considered external fields ($h=0$) and all exchange interactions have the same value ($J_{1}=J_{s}=J_{c}=1$).

In section 2, we show the use of Bogoliubov inequality to obtain an approximate Gibbs free energy and therefore, the magnetization $m_{i}$, the magnetic quadrupole $q$, the internal energy $U$, the specific heat $C$, the entropy $S$ and phases transitions for the cluster of 7 spins described above. In section 3, first we present the phase diagram, the magnetization $M_{T}$ of the nanowire, and its core $m_{c}$ and shell $m_{s}$ for specifics values of the anisotropy $D$. Then, through the Gibbs free energy $G$ and the entropy $S$, the stable, metastable and unstable states are analyzed and used to explain the magnetization behaviors. In section 4, we make the concluding remarks.

\section*{Model and Formulation} \label{MFA} \vspace{0.4cm}
In this section we use the mean field approximation for the mixed spin 1/2 and spin 1 models in a hexagonal nanowire system. By the trial hamiltonian and its respective Gibbs free energy the total Gibbs free energy of the nanowire system was approximated through the Bogoliubov inequality. From this, it was  obtained the phase diagram, magnetizations and entropy.

Considering our nanowire, see Fig-\ref{fig.1}, composed of spin 1/2 Ising in the central column (smaller circles) and spin 1 Blume-Capel in external hexagonal cylinder (bigger circles), the system can be described by the following hamiltonian:
\begin{equation} \label{H}
H = - J_{1} \sum\limits_{\langle i,m \rangle}  S_{i} \sigma_{m}
- J_{c}\sum\limits_{\langle m,n \rangle} \sigma_{m} \sigma_{n}
- J_{s}\sum\limits_{\langle i,j \rangle} S_{i} S_{j}
- D\sum\limits_{i} S_{i}^2,
\end{equation}
where $\sigma_{\alpha}$ or $S_{\alpha}$ is the spin at the lattice site $\alpha$, which takes the values $\sigma = \left\lbrace -\frac{1}{2}, \frac{1}{2} \right\rbrace$ and $S = \left\lbrace -1,0,1 \right\rbrace$. The lower indexes between bracket indicate the sum between the nearest neighbours. $J_{1}$ is the exchange constant between a central spin and its external neighbours, $J_{c}$ is the exchange constant between a central spin and its central neighbours, and $J_{s}$ is the exchange constant between each external hexagonal spin and its external neighbours. From left to right, the Eq. (\ref{H}) represents the interaction between the internal and external spins, between the internal Ising spins and the two terms of external Blume-Capel spins.

For a canonical ensemble the thermal average $\left\langle...\right\rangle$ can be calculated by the trial partition function $Z_{o}$ by:
\begin{equation}
\left\langle ... \right\rangle_{o} = \dfrac{1}{Z_{o}} \mbox{Tr} \left[ (...) \mbox{e}^{-\beta H_{o}} \right]$ and $Z_{o} = \mbox{Tr} \left[ \mbox{e}^{-\beta H_{o}} \right],
\end{equation}
where $\beta$ is the inverse temperature and $\mbox{Tr}$ is the trace for the hamiltonian $H_{o}$.

The total energy can be divided in two parts: a trial unperturbed $H_{o}$ and a perturbed $H^{'}$. By a ratio between the total partition function $Z$ and $Z_{o}$, through the knowledge of $\left\langle e^{(-x)} \right\rangle_{o} \geqslant e^{\left\langle -x \right\rangle_{o}}$ \cite{Strecka}, and the property for classical system $\left[ H_{o},H^{'} \right]=0$ \cite{Kaneyoshi}, we have:
\begin{equation}
\dfrac{Z}{Z_{o}}=\dfrac{ \mbox{Tr} \left[ \mbox{e}^{-\beta H^{'} } \mbox{e}^{-\beta H_{o}} \right]}{ \mbox{Tr} \left[ \mbox{e}^{-\beta H_{o} } \right]} = \left\langle \mbox{e}^{-\beta H^{'}} \right\rangle_{o} \geqslant \mbox{e}^{-\beta \left\langle H^{'} \right\rangle _{o}} 
\end{equation}
and
\begin{equation}
\mbox{ln}Z - \mbox{ln}Z_{o} \geqslant -\beta \left\langle H - H_{o} \right\rangle_{o}.
\end{equation}

From the thermodynamic relationship $G_{i}=-\dfrac{1}{\beta} \mbox{ln} \left[ Z_{i} \right]$, we find the Gibbs-Bogoliubov inequality:
\begin{equation} \label{GBineq}
G \leq \Phi \equiv G_{o} + \langle H - H_{o}\rangle_{o},
\end{equation}
where, $G$ and $H$ represent the true Gibbs free energy and the total hamiltonian of the whole nanowire.

Here, we present a generalization of the mean field approximation for the mixed-spin in a seven site cluster on the nanowire lattice using Bogoliubov inequality. The objective to obtain the mean field approximation on cluster with higher number of spins is to keep the symmetry of the lattice and the core-shell feature. To accomplish this, we use within the mean field approximation a summation over the frontier sites of a cluster of seven sites of the nanowire lattice (see Fig. \ref{fig.1}) and replace an expectation value for all other spins by the mean value. Hence, it follows that the hamiltonian of seven spin is, at the mean field level, given by:
\begin{equation} \label{H7}
H_{7} = -J_{s}\sum\limits_{\vec{\delta_{i}}} S_{i}S_{i+\vec{\delta_{i}}}
- J_{1} \sigma_{0}\sum\limits_{i=1}^{6} S_{i}
- J_{s}\sum\limits_{i=1}^{6} S_{i}\sum\limits_{j=1}^{2} \left\langle S_{i,j} \right\rangle
- D\sum\limits_{i=1}^{6} S_{i}^{2}
- J_{c}\sigma_{0}\sum\limits_{m=1}^{2} \left\langle \sigma_{0,m} \right\rangle. 
\end{equation}
Defining $N'=N/7$, the trial hamiltonian of the whole statistical system is:
\begin{equation} \label{Ho}
H_{o} = \sum\limits_{k=1}^{N'}\left\lbrace
- J_{s}\sum\limits_{\vec{\delta_{i}}} S_{i} S_{i+\vec{\delta_{i}}}
- J_{1} \sigma_{0}\sum\limits_{i=1}^{6} S_{i}
- \gamma_{s}\sum\limits_{i=1}^{6} S_{i}
- \eta\sum\limits_{i=1}^{6} S_{i}^{2}
- \gamma_{c} \sigma_{0} \right\rbrace.
\end{equation}

From the trial partition function $Z_{o}$, we calculated the trial Gibbs free energy $G_{o}$ and therefore, the normalized magnetizations of the core $m_{c}$ and of the shell $m_{s}$, and the quadrupole of the shell $q$:
\begin{equation}
Z_{o} = \sum\limits_{\{\sigma_{m},S_{i}\}}\mbox{exp}(-\beta H_{o})= \prod\limits_{k=1}^{N'}\sum\limits_{\{\sigma_{m},S_{i}\}} \mbox{exp} \left[
\beta \left( J_{s}\sum\limits_{\vec{\delta_{i}}} S_{i} S_{i+\vec{\delta_{i}}}
+ J_{1} \sigma_{0}\sum\limits_{i=1}^{6} S_{i}
+ \gamma_{s}\sum\limits_{i=1}^{6} S_{i}
+ \eta\sum\limits_{i=1}^{6} S_{i}^{2}
+ \gamma_{c} \sigma_{0} \right) \right] \nonumber,
\end{equation}
\begin{equation}
G_{o} = -\dfrac{1}{\beta} \mbox{ln}\left[Z_{o}\right], \label{Go}
\end{equation}
\begin{equation}
m_{i} = -\dfrac{1}{N'}\dfrac{\partial G_{o}}{\partial \gamma_{i}}   \qquad \qquad  \mbox{and} \label{mi} 
\end{equation}
\begin{equation}
q = -\dfrac{1}{N'}\dfrac{\partial G_{o}}{\partial \eta}.   \label{qi}
\end{equation}

Taking $\left\langle S \right\rangle=m_{s}$, $\left\langle S^{2} \right\rangle=q$ and $\left\langle \sigma \right\rangle=m_{c}$, the mean value for the difference between the true and trial hamiltonians is:
\begin{equation} \label{H-Ho}
\left\langle H - H_{o} \right\rangle_{o} = - N'\left( 6 J_{1}  m_{c} m_{s} + J_{c} m_{c}^{2} + 6 J_{s} m_{s}^2 + 6 D q - 6 \gamma_{s}m_{s} - 6 \eta q - \gamma_{c}m_{c} \right),
\end{equation}
which leads in combination with Eq. (\ref{GBineq}) and Eq. (\ref{Go}) to the expression for the variational Gibbs free energy $\Phi$. Through Eq. (\ref{mi}) and Eq. (\ref{qi}) we can minimize $\Phi$ for $\lambda \equiv \{ \gamma_{c}, \gamma_{s}, \eta \}$:
\begin{equation} 
\dfrac{\partial \Phi}{\partial \lambda} = 0 \Leftrightarrow N' \Big(6 J_{1} m_{s} +  2 J_{c} m_{c} -  \gamma_{c} \Big)\dfrac{\partial  m_{c}}{\partial \lambda} 
+N' \Big( 6 J_{1} m_{c} + 12 J_{s}m_{s} - 6\gamma_{s} \Big)\dfrac{\partial  m_{s}}{\partial \lambda} 
 +N' \Big( 6D - 6\eta \Big) \dfrac{\partial q}{\partial \lambda} = 0,
\end{equation}
which are simultaneously satisfied if the variational parameters are $\gamma_{c} = 6J_{1}m_{s} + 2J_{c}m_{c}$ , $\gamma_{s} = J_{1} m_{c} + 2J_{s}m_{s} $ and $\eta=D$. Then, from those effective fields found and through the Eq. (\ref{Go}) and Eq. (\ref{H-Ho}), we have the approximated Gibbs free energy for the seven spin cluster:
\begin{eqnarray}
G &=& -\dfrac{N'}{\beta} \mbox{ln}\left[ \sum\limits_{\{S_{i}, \sigma_{m}\}} \mbox{exp} \left[ \beta \left(
J_{s}\sum\limits_{\vec{\delta_{i}}}S_{i} S_{i+\vec{\delta_{i}}}
+ J_{1} \sigma_{0}\sum\limits_{i=1}^{6}S_{i}
+ \left( J_{1} m_{c} + 2J_{s}m_{s} \right) \sum\limits_{i=1}^{6}S_{i}
+ D\sum\limits_{i=1}^{6}S_{i}^{2}
+ \left( 6J_{1}m_{s} + 2J_{c}m_{c} \right) \sigma_{0} \right) \right] \right]\nonumber \\
& &+ N'\left( 6\,J_{1}{\it m_{c}}\,{\it m_{s}}+{\it J_{c}}\,{{\it m_{c}}}^{2}+6\,{\it J_{s}}\,{{\it m_{s}}}^{2} \right).
\end{eqnarray}
We have sum explicitly over the variable $\sigma_{\alpha}$ and $S_{\alpha}$ and we find the expression for $G$ given by the Eq. (\ref{G}) in section 4.

The second order phase transition occurs at Curie's temperature ($T_{C}$) for the necessary condition $m_{s},m_{c} \rightarrow 0$. The first order phase transition was investigated by the restriction in Gibbs free energy:
\begin{equation} \label{1OPTcond}
G(m_{c},m_{s}) = G(m_{c}^{'},m_{s}^{'})$,   with $m_{c} \neq m_{c}^{'}$ and/or $m_{s} \neq m_{s}^{'}.
\end{equation}

The thermodynamics gives us the entropy $S$, the internal energy $U$ and the specific heat $C$ of the seven spin cluster \cite{Jander_Generalization}:
\begin{equation}
S = - \dfrac{d G}{d T}, \label{Entropia} 
\end{equation}
\begin{equation} 
U = G +TS, \label{EnInt}
\end{equation} 
\begin{equation}
C = T\dfrac{d S}{d T} = T \left[ \dfrac{\partial S}{\partial T} +  \sum\limits_{\alpha= s,c} \dfrac{\partial S}{\partial m_{\alpha}} \dfrac{d m_{\alpha}}{dT} \right]. \label{CalEsp}
\end{equation}

From our choice of the spins distribution in the cluster, the maximum values of the magnetizations $m_{c}$ and $m_{s}$ are 0.5 and 1, respectively. Thus, being the cluster magnetization an average of $m_{c}$ and $m_{s}$, $M_{T}=\dfrac{m_{c}+6m_{s}}{7}$, its maximum value is $\dfrac{6.5}{7} \approx 0.92$.

The numerical results for the critical temperature and phases diagram will be presented in section 3.

\section*{Numerical Results and Diagrams} \label{3}

In this section we present the results obtained by the Gibbs-Bogoliubov inequality and MFA presented in the previous section, for a hexagonal seven spin cluster, Fig. \ref{fig.1}. The phase diagram with the first order, second order and metastable phase transitions was plotted in the $D$ $X$ $T$ plane. The central $m_{c}$, the external $m_{s}$ and the total $M_{T}$ magnetization were calculated for the stable, the metastable and two unstable states. For these four possible states, were calculated the Gibbs free energy, the magnetic quadrupole, the internal energy, the specific heat and the entropy. In special, the unexpected increase in magnetization with temperature and the extinction of the metastable state were investigated. The conclusions are presented in section 4.

\begin{figure}[H]
\centering
   \includegraphics{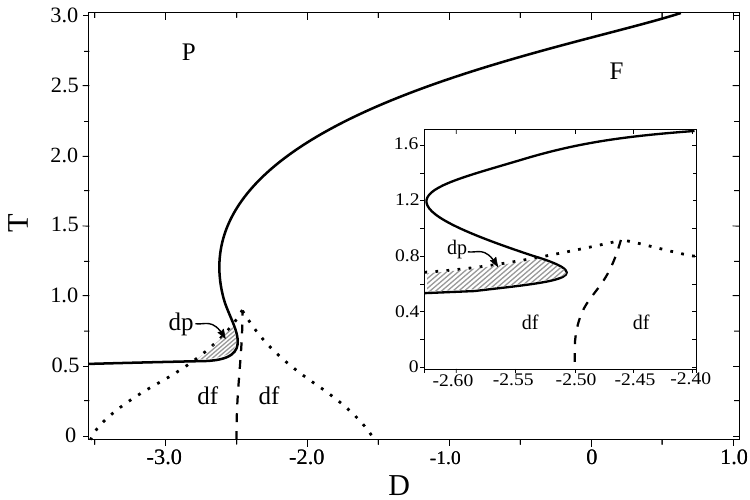}
  \captionof{figure}{Reduced temperature as a function of the anisotropy constant. The solid, the dashed and the dotted curves represent the second order, the first order and the metastable phase transition, respectively. The inset shows in detail the region around the first order phase transition curve for the $D$ axis.}
  \label{diagrama}
\end{figure}

Fig-\ref{diagrama} shows the second order phase transition curve (solid line), which separates the ferromagnetic (F) and paramagnetic (P) phases. Near this curve and in its ferromagnetic region, for any arbitrary $D$ value, the magnetization decrease gradually to zero with the increase of the temperature. The second order phase transition curve presents an asymptotic behavior for $D \rightarrow \infty$ converging for $T\approx4$, and a continuous value, $T=0.5$, for $D \rightarrow - \infty$. These extreme limits correspond to the sum of the expected values, by MFA, for the spin 1/2 Ising system with $z=2$ and spin 1 Blume-Capel system $z=4$. The \textit{inset} in Fig. \ref{diagrama} shows the behavior of the first order phase transition curve (dashed line), which has $D=-2.50$ for $T \rightarrow 0$ and its highest temperature value is 0.90 for $D=-2.46$. For any arbitrary value of $D$ between these limits, with the crossing of the first order phase transition curve by temperature increase, both the core and the shell suffer an abrupt transition to lower values of magnetization, but neither go to zero. The dotted curve in Fig. \ref{diagrama} limits the region in $T$ $x$ $D$ plan with the presence of a metastable state. Its superior limit is the same of the first order phase transition curve and its bottom limits are $D$ equal $-3.52$ and $-1.53$ for $T \rightarrow 0$. The second order phase transition curve distinguishes the dense paramagnetic phase dp (hatched region) of the dense ferromagnetic phase df \cite{ekiz_dfdp}, both in the metastable region. The absence of the metastable state out of the metastable region is due a soft overlap of this state with an unstable state, resulting in the extinction of both. In this work, this kind of transition will be called \textit{metastable transition}. This transition will better explained below.

\begin{figure}[H]
  \centering
  \includegraphics{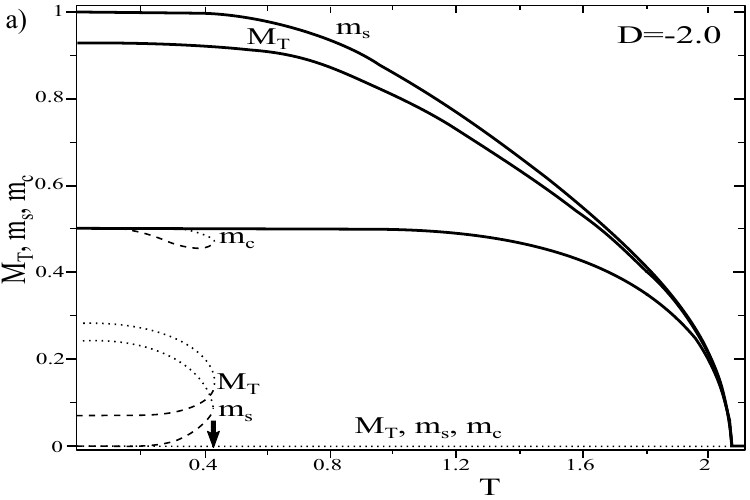}
  \includegraphics{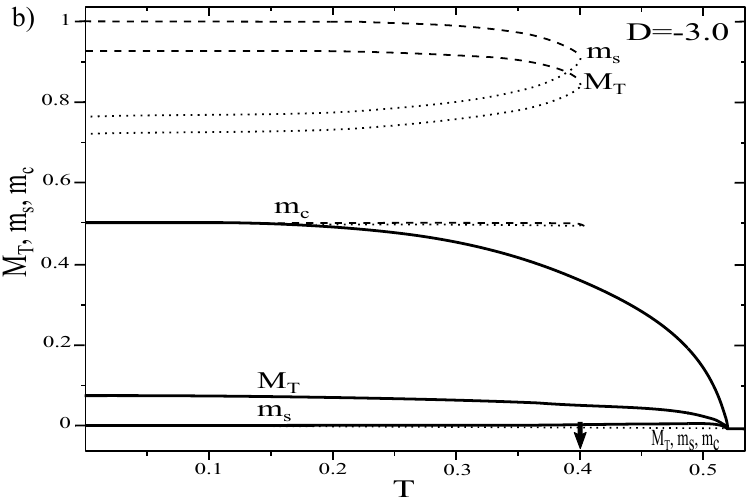}
  \captionof{figure}{Magnetization curves of the nanowire $M_{T}$, of the core $m_{c}$, and of the shell $m_{s}$ as function of the temperature, for (a) $D=-2$ and (b) $D=-3$ (b). The solid, the dashed, and the two dotted lines represent the stable, the metastable and the two unstable states of the system, respectively.}
  \label{mxT-2-3}
\end{figure}

To study the magnetization behavior of the system, we fixed $D$ while increasing the temperature from 0 to the respective $T_{C}$. The chosen D values were $-2$, $-3$ and $-2.48$. The three values give a metastable transition, but just $D=-2.48$ shows a first order phase transition. For the three anisotropy values are shown the core $m_{c}$,  the shell $m_{s}$ and the total $M_{T}$ magnetizations for the four possible states, a stable, a metastable and two unstable. For $D=-2$, Fig. \ref{mxT-2-3} (a), the stable magnetization (solid line) of the system behaves like a simple Ising ferromagnetic \cite{Jander_Generalization}, having its maximum values as $T \rightarrow 0$ and decreasing to zero as $T \rightarrow T_{C}=2.08$. The metastable magnetization (dashed line) is observed with constant value for $T < 0.2$, followed by a small decrease in $m_{c}$, and an increase in $m_{s}$ and $M_{T}$ up to $T = 0.4$. Two unstable magnetizations are observed for $D=-2$, one with value always equal zero (bottom dotted line) and the other with value not zero (superior dotted line). The first one meets with the stable magnetization in the $T_{C}$ point, while the second one meets with the metastable magnetization before the second order phase transition, in $T=0.42$, resulting in a metastable transition. A black arrow in Fig. \ref{mxT-2-3} (a) indicates in the temperature axis the metastable transition point. For $D=-3$, Fig. \ref{mxT-2-3} (b), we observe the presence of stable (solid), metastable (dashed) and two unstable (dotted lines) magnetizations. Different of the $D=-2$, the metastable transition occurs for values of the magnetization higher than those of the stable state, meanly for the shell spins. This transition occurs at $T=0.40$. A little increase in $m_{s}$ is observed before the second order phase transition, which occurs at $T_{C}=0.52$.

\begin{figure}[H]
  \centering
  \includegraphics{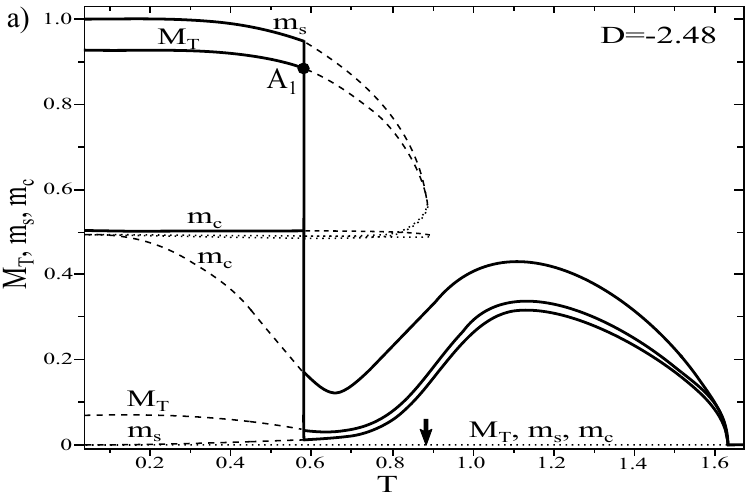}
  \includegraphics{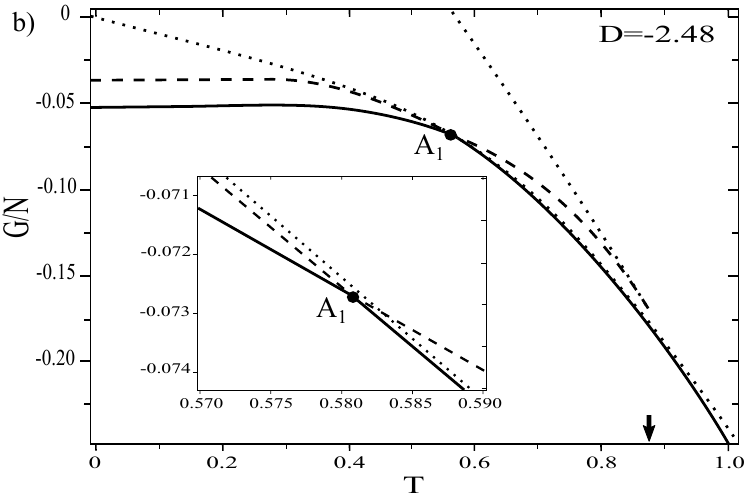}
  \captionof{figure}{a) Magnetization curves of the nanowire $M_{T}$, of the core $m_{c}$ and of the shell $m_{s}$ as function of the temperature, for $D=-2.48$. b) Gibbs free energy as function of the temperature for $D=-2.48$. The solid, the dashed, and the dotted lines represent the stable, the metastable and the two unstable states of the system, respectively. The inset evidences the change in direction between the stable and the metastable Gibbs free energy curves in the first order phase transition point, $A_{1}$.}
  \label{mxT-2.48}
\end{figure}

In the \textit{inset} in Fig. \ref{diagrama}, one sees that for $D=-2.48$ the system suffers a first order phase transition, at $T=0.58$, a metastable transition, at $T=0.88$, and a second order phase transition, at $T_{C}=1.63$. In the magnetizations curves for $D=-2.48$, Fig. \ref{mxT-2.48} (a), the stable magnetization (solid line) has the maximum value for $T\rightarrow0$. For $T<0.58$, the unstable magnetization with $m_{c}=m_{s}=M_{T} \approx 0.5$ (superior dotted line) has higher values than the metastable magnetization one (dashed line), as has been seen for $D=-2$ (Fig. \ref{mxT-2-3} (a)). Increasing the temperature, in $T=0.58$, the system presents a first order phase transition with its stable and metastable states, which causes an abrupt decrease in the stable magnetization. This transition is observed both in $m_{c}$ and in $m_{s}$, although $m_{c}$ doesn't have anisotropic interaction. For $T > 0.58$, the possibles states are similar as in the $D=-3$ situation, the metastable magnetization with highest value and presenting a metastable transition, which occurs at $T=0.88$. As in Fig. \ref{mxT-2-3} (b), the magnetizations of the stable state show an increase with the increase of temperature before $T_{C}$. This will be better observed through the Gibbs free energy states paths, Fig. \ref{G(mc,me)}. Fig. \ref{mxT-2.48} (b) shows the Gibbs free energy as a function of temperature, for $D=-2.48$, of the stable (solid line), of the metastable (dashed line) and of the unstable states (dotted lines). The \textit{inset} evidences the first order phase transition point, $A_{1}$, where the stable and the metastable curves change their directions. In the metastable transition, identified by a black arrow in the temperature axis, the metastable and unstable states show the same Gibbs free energy value.

\begin{figure}[H]
  \centering
  \includegraphics{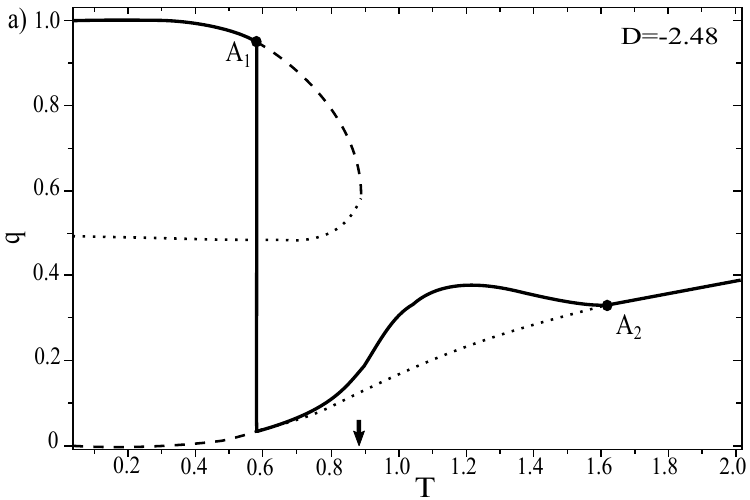}
  \includegraphics{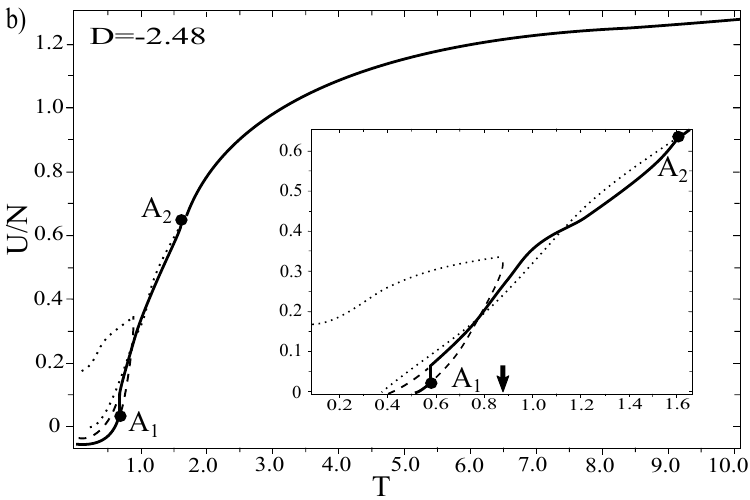}
  \captionof{figure}{Magnetic quadrupole $q$ of the shell spins (figure a) and internal energy per spin $U/N$ of the nanowire (figure b), as function of the temperature, for $D=-2.48$. The solid, the dashed and the two dotted lines represent the stable, the metastable and the two unstable states, respectively. The inset in b) evidences the discontinuity in the stable curve of the internal energy at the first order phase transition point, $A_{1}$, and the metastable transition point in the temperature axis, black arrow}
  \label{qU}
\end{figure}

Fig. \ref{qU} a) shows the curves of the shell's magnetic quadrupole $q$ by temperature, of the four possible states reached for $D=-2.48$, one stable, one metastable and two unstable. The first order phase transition point, $A_{1}$, the second order phase transition point, $A_{2}$, and the metastable transition, black arrow, are indicated in the graphic. For $T \rightarrow \infty$, $q$ converges to an asymptotic value of $q=0.67$. In the internal energy per spin $U/N$, Fig. \ref{qU} b), for $T \rightarrow \infty$, the internal energy converges to the asymptotic value of $U/N = 1.42$. The relationship between these limit values becomes clear by the definition of $U/N = \left\langle H \right\rangle$, see Eq. (\ref{H}), and by the approximations mean values made in Section-2, between the Eq. (\ref{qi}) and Eq. (\ref{H-Ho}). Then, straightforwardly, we have $\dfrac{1}{N}U(T \rightarrow \infty) = - \dfrac{6D}{7}q(T \rightarrow \infty)$.

\begin{figure}[H]
  \centering
  \includegraphics{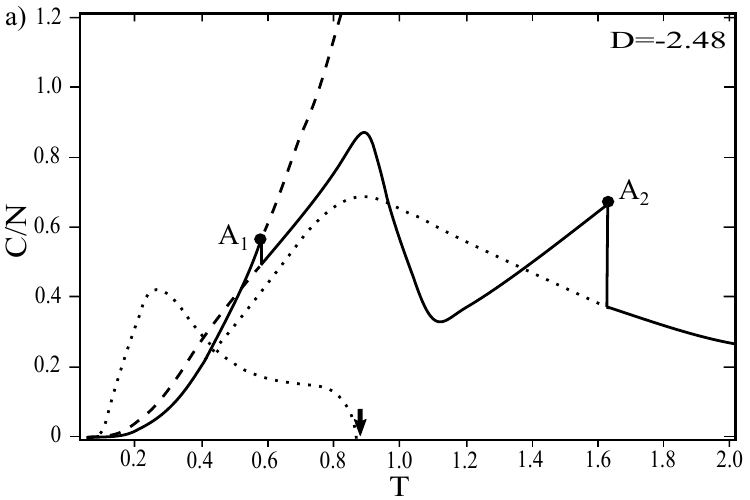}
  \includegraphics{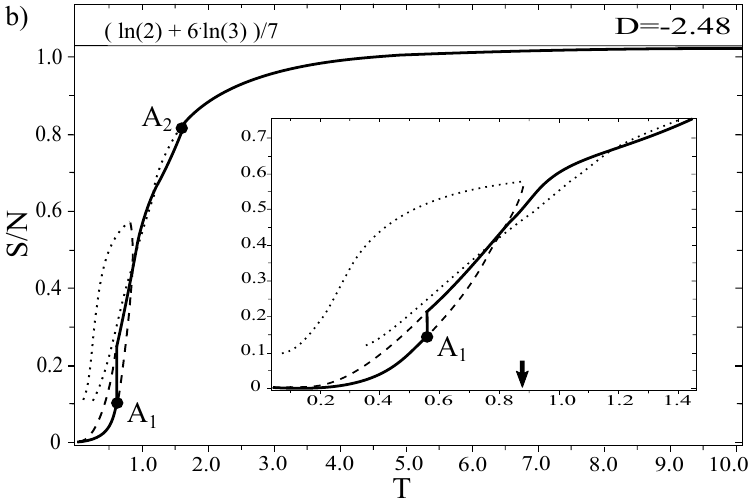}
  \captionof{figure}{Specific heat per spin $C/N$ (figure a) and entropy per spin $S/N$ (figure b) of the nanowire, as function of the temperature, for $D=-2.48$. The solid, the dashed and the two dotted lines represent the stable, the metastable and the two unstable states, respectively. The inset in b) evidences the discontinuity in the stable curve of the entropy at the first order phase transition point, $A_{1}$, and the metastable transition point in the temperature axis, black arrow}
  \label{CS}
\end{figure}

Fig. \ref{CS} a) shows the states' curves of the specific heat per spin $C/N$, by the temperature, for $D=-2.48$. In the metastable transition, pointed by the black arrow, the metastable and the unstable curves diverge to infinity and minus infinity. These behaviors are expected by $\dfrac{\partial m_{i}}{\partial T}$ component of Eq. (\ref{CalEsp}) and magnetizations curves behavior near this transition, Fig. \ref{mxT-2.48} a). For $T \rightarrow \infty$, the entropy per spin $S/N$, Fig. \ref{CS} b), converges to the expected value for a system with $\dfrac{N}{7}$ spins type Ising and $\dfrac{6N}{7}$ spins type Blume-Capel, Fig. \ref{fig.1}. A discontinuous in the second order phase transition point $A_{2}$ is observed just in the specific heat curve, while a discontinuous in the first order phase transition point $A_{1}$ is observed in the entropy curve.

\begin{figure}[H]
  \centering
  \includegraphics{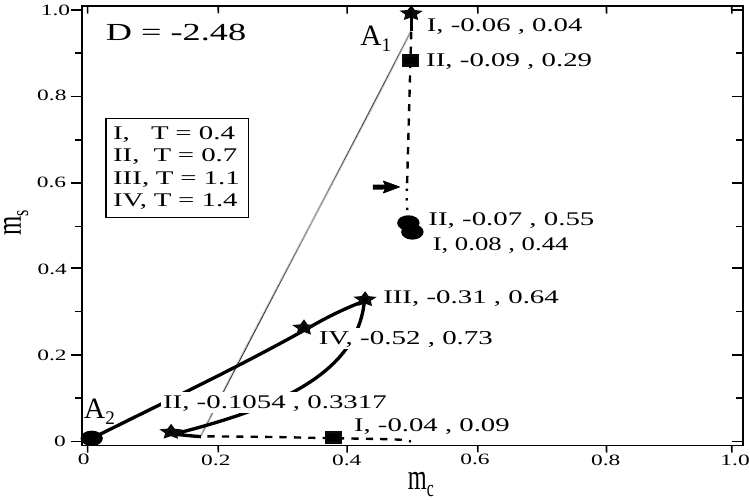}
  \captionof{figure}{$m_{c}$ $x$ $m_{s}$ plane with the paths of the stable (solid), the metastable (dashed) and the unstable (dotted lines) states. Specific values of the stable, the metastable and the two unstable states for $T=0.4$, 0.7, 1.1 and 1.4 are indicated by the star, the square and the full circles, respectively. For the unstable state with zero magnetization, the $G/N$ and the $S/N$ are -0.04 , 0.12 for $T=0.4$, -0.1053 , 0.3311 for $T=0.7$, -0.30 , 0.63 for $T=1.1$ and -0.51 , 0.76 for $T=1.4$.}
  \label{G(mc,me)}
\end{figure}

Fig. \ref{G(mc,me)} shows the Gibbs free energy paths of each state, in the $m_{c}$ $\times$ $m_{s}$ plane, for $0 < T < T_{C}$ and $D=-2.48$. The solid, the dashed and the dotted lines represent the paths \cite{caminhos_ekiz} of the stable, the metastable and the unstable states. The star, the square and the full circles represent the stable, the metastable and the two unstable states in four specific temperatures, $T_{I}=0.4$, $T_{II}=0.7$, $T_{III}=1.1$ and $T_{IV}=1.4$. The pair of numbers inserted in the graphics are the Gibbs free energy per spin $G/N$ and entropy per spin $S/N$ values, for each thermodynamic state of a given temperature. The first order phase transition point, $A_{1}$, the second order phase transition point, $A_{2}$, and the metastable transition, black arrow, are represented in the figure.
As observed above, Fig. \ref{G(mc,me)} presents, for $T=0.4$, the magnetizations $m_{c}=0.50$ and $m_{s}=0.99$ for the stable state, $m_{c}=0.38$ and $m_{s} \approx 0$ for the metastable state,  $m_{c}=0.50$ and $m_{s}0.49$ for an unstable state and $m_{c}=m_{s}=0$ for the other unstable state. After the first order phase transition, by the motion of the star from $T=0.7$ to and $T=1.1$, we observe the increase of the magnetization of the stable state, mainly in the external spins ($m_{s}$). This increase occurs toward  $m_{c} \approx 0.5$ and $m_{c} \approx 0.5$. Simultaneously, the metastable state (square) decrease its external magnetization towards  the same point, reaching the unstable state in $m_{c}=0.49$ and $m_{s}=0.59$ for $T=0.88$. After this metastable transition, for $T=1.1$ and $T=1.4$, we observe the  magnetization of the stable state decreasing towards the unstable state in zero magnetization.

\section*{Final Comments}

We presented an analysis, through the Gibbs-Bogoliubov inequality and MFA, of the thermodynamics of the states of a system composed of Ising and Blume-Capel core-shell spins. Comparing the anisotropy value above and below the first order phase transition limits, were observe inversions in the magnetization behavior between stable, metastable and unstable states. For the anisotropy below this limit, with $T << T_{C}$, the system does not present magnetic orientation in its external Blume-Capel spins.
The observed presence of the first order phase transition both in the Ising and in the Blume-Capel spins, is a clear evidence of a high influence between the spins of the core and of the shell in the ordered phase.

As expected by the thermodynamic relationship between the Gibbs free energy, entropy and specific heat, $S \propto \dfrac{\partial G}{\partial T}$ and $C \propto \dfrac{\partial^{2} G}{\partial T^{2}}$, a discontinuity in the first order and in the second order phase transition points were observed in the entropy and specific heat curves, respectively. The discontinuity in the specific heat curve in the metastable transition point, indicates a second order phase transition behavior of this transition.
An increase in magnetization with the increase of the temperature, was observed for an anisotropy between the limits of first order phase transition and right below it. This behavior was more evident in the external Blume-Capel spins. Analyzing the entropy values of the two unstable states presented, before and after the metastable transition, we observed a tendency of the stable and the metastable states to approximate to the unstable state with highest entropy value.
The use of the Bogoliubov inequality in the MFA gave us a Gibbs free energy nearer of its real value than that obtained for the seven site cluster, $G_{o}$. It ensured us a more precise curve of internal energy and first order phase transition, except when $m \rightarrow 0$.

\section*{Appendix}

In this section we present the explicit expression for $G$ without external fields.

\begin{eqnarray} 
\label{G}
G=-N'k_{{B}}T\ln \Big(Z_m\Big)+ N'\left( 6\,J_{1}{\it m_{c}}\,{\it m_{s}}+{\it J_{c}}\,{{\it m_{c}}}^{2}+6\,{\it J_{s}}\,{{\it m_{s}}}^{2} \right)
\end{eqnarray}

\begin{eqnarray}
Z_{m}&=&  \big[ 1+6\,{{\rm e}^{\beta\, \left( -J_{1}{\it m_{c}}-2\,{\it J_{s}}\,{\it m_{s}}
+{\it D} \right) }}+6\,{{\rm e}^{\beta\, \left( J_{1}{\it m_{c}}+2
\,{\it J_{s}}\,{\it m_{s}}+{\it D} \right) }}+24\,{{\rm e}^{4\,
\beta\,{\it D}}}+12\,{{\rm e}^{\beta\, \left( {\it J_{s}}+4\,{\it D}
 \right) }}+18\,{{\rm e}^{2\,\beta\,{\it D}}} \nonumber \\
 & &+12\,{{\rm e}^{\beta\, \left( -3\,{\it J_{s}}+4\,{\it D} \right) }}+12\,{{\rm e}^{\beta\,
 \left( -{\it J_{s}}+2\,{\it D} \right) }}+12\,{{\rm e}^{\beta\, \left( 
-{\it J_{s}}+4\,{\it D} \right) }}+6\,{{\rm e}^{\beta\, \left( -5\,J_{1}{
\it m_{c}}-10\,{\it J_{s}}\,{\it m_{s}}+4\,{\it J_{s}}+5\,{\it D}
 \right) }} \nonumber \\
 & &+6\,{{\rm e}^{\beta\, \left( -4\,J_{1}{\it m_{c}}-8\,{\it J_{s}}\,{
\it m_{s}}+3\,{\it J_{s}}+4\,{\it D} \right) }}+12\,{{\rm e}^{
\beta\, \left( -3\,J_{1}{\it m_{c}}-6\,{\it J_{s}}\,{\it m_{s}}+2\,{
\it J_{s}}+5\,{\it D} \right) }}+6\,{{\rm e}^{\beta\, \left( -3\,J_{1}{\it 
m_{c}}-6\,{\it J_{s}}\,{\it m_{s}}+2\,{\it J_{s}}+3\,{\it D} \right) 
}} \nonumber \\
& &+18\,{{\rm e}^{\beta\, \left( -3\,J_{1}{\it m_{c}}-6\,{\it J_{s}}\,{\it m_{s}}
+5\,{\it D} \right) }}+12\,{{\rm e}^{\beta\, \left( -J_{1}{\it 
m_{c}}-2\,{\it J_{s}}\,{\it m_{s}}+2\,{\it J_{s}}+5\,{\it D} \right) }}+
12\,{{\rm e}^{\beta\, \left( -2\,J_{1}{\it m_{c}}-4\,{\it J_{s}}\,{\it m_{s}}
+{\it J_{s}}+4\,{\it D} \right) }} \nonumber \\
& &+12\,{{\rm e}^{\beta\, \left( -3\,J_{1}{\it m_{c}}-6\,{\it J_{s}}\,{\it m_{s}}+{\it J_{s}}+3\,{\it D} \right) }}+6\,{{\rm e}^{\beta\, \left( -2\,J_{1}{\it m_{c}}-4\,{\it J_{s}}\,{
\it m_{s}}+{\it J_{s}}+2\,{\it D} \right) }}+12\,{{\rm e}^{
\beta\, \left( -J_{1}{\it m_{c}}-2\,{\it J_{s}}\,{\it m_{s}}+{\it J_{s}}+3\,{
\it D} \right) }} \nonumber \\
& &+12\,{{\rm e}^{\beta\, \left( -2\,J_{1}{\it m_{c}}-4\,{\it 
J_{s}}\,{\it m_{s}}-{\it J_{s}}+4\,{\it D} \right) }}+24\,{{\rm e}
^{\beta\, \left( -J_{1}{\it m_{c}}-2\,{\it J_{s}}\,{\it m_{s}}-2\,{\it J_{s}}
+5\,{\it D} \right) }}+18\,{{\rm e}^{\beta\, \left( -J_{1}{\it m_{c}}-2\,{
\it J_{s}}\,{\it m_{s}}+3\,{\it D} \right) }} \nonumber \\
& &+18\,{{\rm e}^{\beta\, \left( -J_{1}{\it m_{c}}-2\,{\it J_{s}}\,{\it m_{s}}+5\,{\it D} \right) }}+12\,{{\rm e}^{\beta\, \left( J_{1}{\it m_{c}}+2\,{\it J_{s}}\,{\it 
m_{s}}+2\,{\it J_{s}}+5\,{\it D} \right) }}+24\,{{\rm e}^{\beta\,
 \left( -J_{1}{\it m_{c}}-2\,{\it J_{s}}\,{\it m_{s}}-{\it J_{s}}+3\,{\it D}
 \right) }} \nonumber \\
 & &+12\,{{\rm e}^{\beta\, \left( J_{1}{\it m_{c}}+2\,{\it J_{s}}\,{\it 
m_{s}}+{\it J_{s}}+3\,{\it D} \right) }}+18\,{{\rm e}^{\beta\,
 \left( J_{1}{\it m_{c}}+2\,{\it J_{s}}\,{\it m_{s}}+3\,{\it D} \right) }
}+6\,{{\rm e}^{\beta\, \left( -J_{1}{\it m_{c}}-2\,{\it J_{s}}\,{\it m_{s}}
-2\,{\it J_{s}}+3\,{\it D} \right) }} \nonumber \\
& &+6\,{{\rm e}^{\beta\, \left( -J_{1}{
\it m_{c}}-2\,{\it J_{s}}\,{\it m_{s}}-4\,{\it J_{s}}+5\,{\it D}
 \right) }}+24\,{{\rm e}^{\beta\, \left( J_{1}{\it m_{c}}+2\,{\it J_{s}}\,{\it 
m_{s}}-2\,{\it J_{s}}+5\,{\it D} \right) }}+24\,{{\rm e}^{\beta\,
 \left( J_{1}{\it m_{c}}+2\,{\it J_{s}}\,{\it m_{s}}-{\it J_{s}}+3\,{\it D}
 \right) }} \nonumber \\
 & &+18\,{{\rm e}^{\beta\, \left( J_{1}{\it m_{c}}+2\,{\it J_{s}}\,{\it 
m_{s}}+5\,{\it D} \right) }}+12\,{{\rm e}^{\beta\, \left( 2\,J_{1}{
\it m_{c}}+4\,{\it J_{s}}\,{\it m_{s}}+{\it J_{s}}+4\,{\it D}
 \right) }}+12\,{{\rm e}^{\beta\, \left( 3\,J_{1}{\it m_{c}}+6\,{\it J_{s}}\,{
\it m_{s}}+2\,{\it J_{s}}+5\,{\it D} \right) }} \nonumber \\
& &+6\,{{\rm e}^{\beta\, \left( 2\,J_{1}{\it m_{c}}+4\,{\it J_{s}}\,{\it m_{s}}
+{\it J_{s}}+2\,{\it D} \right) }}+6\,{{\rm e}^{\beta\, \left( J_{1}{\it m_{c}}+2\,{\it 
J_{s}}\,{\it m_{s}}-2\,{\it J_{s}}+3\,{\it D} \right) }}+6\,{{\rm e}^
{\beta\, \left( 3\,J_{1}{\it m_{c}}+6\,{\it J_{s}}\,{\it m_{s}}+2\,{
\it J_{s}}+3\,{\it D} \right) }} \nonumber \\
& &+12\,{{\rm e}^{\beta\, \left( 2\,J_{1}{\it 
m_{c}}+4\,{\it J_{s}}\,{\it m_{s}}-{\it J_{s}}+4\,{\it D} \right) }}+
12\,{{\rm e}^{\beta\, \left( 3\,J_{1}{\it m_{c}}+6\,{\it J_{s}}\,{\it m_{s}}
+{\it J_{s}}+3\,{\it D} \right) }} +6\,{{\rm e}^{\beta\, \left( 4\,J_{1}{\it m_{c}}
+8\,{\it J_{s}}\,{\it m_{s}}+3\,{\it J_{s}}+4\,{\it D} \right) }}\nonumber \\
& &+6\,{{\rm e}^{\beta\, \left( J_{1}{\it m_{c}}+2\,{\it J_{s}}\,{\it m_{s}
}-4\,{\it J_{s}}+5\,{\it D} \right) }}+18\,{{\rm e}^{\beta\,
 \left( 3\,J_{1}{\it m_{c}}+6\,{\it J_{s}}\,{\it m_{s}}+5\,{\it D}
 \right) }}+6\,{{\rm e}^{\beta\, \left( 5\,J_{1}{\it m_{c}}+10\,{\it J_{s}}\,{
\it m_{s}}+4\,{\it J_{s}}+5\,{\it D} \right) }} \nonumber \\
& &+6\,{{\rm e}^{2\,\beta\, \left( {\it J_{s}}+2\,{\it D} \right) }}+24\,{{\rm e}^{2\,
\beta\, \left( -{\it J_{s}}+2\,{\it D} \right) }}+9\,{{\rm e}^{2\,\beta
\, \left( -2\,J_{1}{\it m_{c}}-4\,{\it J_{s}}\,{\it m_{s}}+{\it J_{s}}+2\,
{\it D} \right) }}+6\,{{\rm e}^{2\,\beta\, \left( -J_{1}{\it m_{c}}-2\,{\it 
J_{s}}\,{\it m_{s}}+{\it J_{s}}+2\,{\it D} \right) }} \nonumber \\
& &+24\,{{\rm e}^{2\,\beta\, \left( -J_{1}{\it m_{c}}-2\,{\it J_{s}}\,{\it m_{s}}+2\,{\it D} \right) }}+9\,{{\rm e}^{2\,\beta\, \left( -J_{1}{\it m_{c}}-2\,{\it J_{s}}\,{
\it m_{s}}+{\it D} \right) }}+2\,{{\rm e}^{3\,\beta\, \left( -J_{1}
{\it m_{c}}-2\,{\it J_{s}}\,{\it m_{s}}+{\it D} \right) }}+6\,{
{\rm e}^{2\,\beta\, \left( -J_{1}{\it m_{c}}-2\,{\it J_{s}}\,{\it m_{s}}-{
\it J_{s}}+2\,{\it D} \right) }} \nonumber \\
& &+6\,{{\rm e}^{2\,\beta\, \left( J_{1}{\it m_{c}}+2\,{\it J_{s}}\,{\it m_{s}}+{\it J_{s}}+2\,{\it D} \right) }}+24\,{{\rm e}^{2\,\beta\, \left( J_{1}{\it m_{c}}+2\,{\it J_{s}}\,{\it m_{s}}+2\,{\it D} \right) }}+9\,{{\rm e}^{2\,\beta\, \left( J_{1}{\it m_{c}}+2\,{\it J_{s}}\,{\it m_{s}}+{\it D} \right) }}+6\,{{\rm e}^{2\,\beta\,
 \left( J_{1}{\it m_{c}}+2\,{\it J_{s}}\,{\it m_{s}}-{\it J_{s}}+2\,{\it D}
 \right) }} \nonumber \\
 & &+2\,{{\rm e}^{3\,\beta\, \left( J_{1}{\it m_{c}}+2\,{\it J_{s}}\,{
\it m_{s}}+{\it D} \right) }}+9\,{{\rm e}^{2\,\beta\, \left( 2
\,J_{1}{\it m_{c}}+4\,{\it J_{s}}\,{\it m_{s}}+{\it J_{s}}+2\,{\it D}
 \right) }}+{{\rm e}^{6\,\beta\, \left( -J_{1}{\it m_{c}}-2\,{\it J_{s}}\,{\it 
m_{s}}+{\it J_{s}}+{\it D} \right) }}+6\,{{\rm e}^{2\,\beta\,
 \left( -2\,J_{1}{\it m_{c}}-4\,{\it J_{s}}\,{\it m_{s}}+{\it J_{s}}+3\,{
\it D} \right) }} \nonumber \\
& &+6\,{{\rm e}^{2\,\beta\, \left( -J_{1}{\it m_{c}}-2\,{\it J_{s}}\,{\it m_{s}}+{\it J_{s}}+3\,{\it D} \right) }}+9\,{{\rm e}^{2\,\beta\, \left( -J_{1}{\it m_{c}}-2\,{\it J_{s}}\,{\it m_{s}}-{\it J_{s}}+3\,{\it D} \right) }}+6\,{{\rm e}^{2\,\beta\, \left( {\it J_{s}}+3\,{\it D} \right) }}+12\,{{\rm e}^{2\,\beta\, \left( -{\it J_{s}}+3\,{\it D}
 \right) }} \nonumber \\
 & &+6\,{{\rm e}^{2\,\beta\, \left( J_{1}{\it m_{c}}+2\,{\it J_{s}}\,{
\it m_{s}}+{\it J_{s}}+3\,{\it D} \right) }}+2\,{{\rm e}^{6\,\beta
\, \left( -{\it J_{s}}+{\it D} \right) }}+9\,{{\rm e}^{2\,\beta\,
 \left( J_{1}{\it m_{c}}+2\,{\it J_{s}}\,{\it m_{s}}-{\it J_{s}}+3\,{\it D}
 \right) }}+6\,{{\rm e}^{2\,\beta\, \left( 2\,J_{1}{\it m_{c}}+4\,{\it J_{s}}\,{
\it m_{s}}+{\it J_{s}}+3\,{\it D} \right) }} \nonumber \\
& &+{{\rm e}^{6\,\beta\, \left( J_{1}{\it m_{c}}+2\,{\it J_{s}}\,{\it m_{s}}+{\it J_{s}}+{\it D} \right) }} \big]   \left[ {{\rm e}^{-1/2\,\beta\, \left( 6\,J_{1}{\it m_{s}
}+2\,{\it J_{c}}\,{\it m_{c}} \right) }}+{{\rm e}^{1/2\,\beta\,
 \left( 6\,J_{1}{\it m_{s}}+2\,{\it J_{c}}\,{\it m_{c}} \right) }} \right]
\end{eqnarray}

\pagebreak
 

\begin{thebibliography}{18}

\bibitem{ExpNanowire1} J. H. Gao, Q. F. Zhan, W. He, D. L. Sun,Z. H. Cheng, Appl. Phys. Lett. 86 (2005) 232506.
\bibitem{ExpNanowire2} Y. Peng, T. Cullis, G. M{\"o}bus, X. Xu, I. Beverley, Nanotechnology 18(2007) 485704.
\bibitem{ExpNanowire3}  HN Hu, HY Chen, SY Yu, LJ Chen, JL Chen, GH Wu, J. Magn. Magn. Mater. 299 (2006) 170.
\bibitem{ExpNanowireMEDICINAL}  A. Fert and L. Piraux, J. Magn. Magn. Mater, 200 (1999) 338.
\bibitem{ExpNanowireMEDICINAL2} J-E Wegrowe, D. Kelly, Y. Jaccard, Ph Guittienne, J-Ph Ansermet, Europhys. Lett. 45 (1999) 626.
\bibitem{EXpNwZnOSensor}  Z. Fan and J. G. Lu, IJHSES, 16 (2006) 883.
\bibitem{MixedNwEFT} W. Jiang, F. Zhang, X. Li, H. Y. Guan, A. B. Guo, Z. Wang, Physica E 47 (2013) 95.
\bibitem{MixedNwEFT2} E. Kantar, Y. Kocakaplan, Solid State Commun. 117 (2014) 1.
\bibitem{MixedNwEFT3} E. Kantar, Y. Kocakaplan, J. Magn. Magn. Mater. 393 (2015) 574.
\bibitem{MixedNtEFT} O. Canko, A. Erdin{\c{c}}, F. Ta{\c{s}}k{\i}n, A. F. Y{\i}ld{\i}r{\i}m, J. Magn. Magn. Mater. 324 (2012) 508.
\bibitem{MixedNwEFTMonteCarlo} M. Boughrara, M. Kerouad, A. Zaim, J. Magn. Magn. Mater. 360 (2014) 222.
\bibitem{MixedNwMonteCarlo} A. Feraoun, A. Zaim, M. Kerouad, M, Physica B 445 (2014) 74.
\bibitem{MixedNwMonteCarlo2} R. Masrour, A. Jabar, A. Benyoussef, M.  Hamedoun, L. Bahmad, Physica B 472 (2015) 19.
\bibitem{J1}  N. Bogoliubov, J. Phys, 11 (1947) 23.
\bibitem{J2}  R. P. Feynman, Phys. Rev. 97 (1955) 660.
\bibitem{J3}  H. Falk, Am. J. Phys. 38 (1970) 858.
\bibitem{J7}  A. L. Kuzemsky, Phys. Part. Nucl. 40 (2009) 949.
\bibitem{J8}  A. L. Kuzemsky, Int. J. Mod. Phys. B, 29 (2015) 1530010.
\bibitem{J9}  L. A. Maksimov and A. L. Kuzemsky, Phys. MET. Metallogr. (USRR) 31 (1971) 5.
\bibitem{J10} J. P. Santos, F. C. S{\'a} Barreto, D. S. Rosa, J. Magn. Magn. Mater. 423 (2017) 175.
\bibitem{Weiss1}  P. Weiss, Compt.Rend. 143 (1906) 1137.
\bibitem{Weiss2}  P. Weiss, J. Phys. (Paris), 6 (1907) 666.
\bibitem{Correlation_Kaneyoshi}  T. Kaneyoshi. Phys. Status Solidi B 242 (2005) 2938.
\bibitem{Jander_Generalization}  J. P. Santos. Braz. J. Phys. 47 (2017) 122.
\bibitem{ekiz_dfdp}  C. Ekiz, M. Keskin, Phys. Rev. B 66 (2002) 054105.
\bibitem{caminhos_ekiz}  C. Ekiz, M. Keskin, O. Yal{\c{c}}{\i}n, Physica A 293 (2001) 215.
\bibitem{Strecka}  J. Strecka and M. JaSCur, Acta Phys. Slov. 65 (2015) 235.
\bibitem{Kaneyoshi}  T. Kaneyoshi, Acta Phys. Pol. A 83 (1993) 703.

\end{thebibliography}

\end{document}